\documentclass[a4paper,11pt]{article}
\usepackage{pos}

\title{Status of the VERITAS Stellar Intensity Interferometry (VSII) System}
 \ShortTitle{VERITAS Stellar Intensity Interferometer}

\author*[a]{David Kieda}
\author[a,b]{Jonathan Davis}
\author[a]{Tugdual LeBohec}
\author[c]{Mike Lisa}
\author[a,d]{Nolan K. Matthews}
\affiliation[a]{University of Utah,\\
  Department of Physics and Astronomy, 115 S 1400 E \#201, Salt Lake City, Utah, USA }
   \affiliation[b]{current address: Cornell University,\\
  Department of Astronomy, Ithaca, NY,  USA }
    \affiliation[c]{The Ohio State University,\\
  Department of Physics, Columbus, OH, USA }
    \affiliation[d]{current address: Universit\'{e} Côte d'Azur, CNRS, INPHYNI, France}  
  
\forColl{VERITAS} % W/O "Collaboration"

\emailAdd{dave.kieda@utah.edu}

\abstract{The VERITAS Imaging Air Cherenkov Telescope array (IACT) was augmented in 2019 with high-speed focal plane electronics to allow the use of VERITAS for Stellar Intensity Interferometry (SII) observations. Since that time, several improvements have been implemented to increase the sensitivity of the VERITAS Stellar Intensity Interferometer (VSII) and increase the speed of nightly data processing. This poster will describe the use of IACT arrays for performing ultra-high resolution (sub-milliarcsecond) astronomical observations at short visible wavelengths. The poster presentation will include a description of the VERITAS-SII focal plane, data acquisition, and data analysis systems. The poster concludes with a description of plans for future upgrades of the VSII instrument.}

\FullConference{37$^{\rm{th}}$ International Cosmic Ray Conference (ICRC 2021)\\
		July 12th -- 23rd, 2021\\
		Online -- Berlin, Germany}

\begin{document}
\maketitle

\section{Introduction}
The Stellar Intensity interferometry (SII) technique measures correlated fluctuations in light intensity between spatially separated telescopes.   The VERITAS Stellar Intensity Interferometer (VSII)  is currently the world's most sensitive SII astronomical observatory. VSII is implemented through an instrumentation augmentation of the existing VERITAS gamma-ray observatory. VSII operates for 5-10 days/month during bright moon periods when VERITAS gamma-ray observations are severely limited by the intensity of scattered moonlight. This paper describes the VSII instrumentation, including recent upgrades and improvements in the VSII observatory.  An accompanying paper at this conference  \cite{Kieda2021} describes a comprehensive VSII survey of northern sky bright stars that is currently in progress. This survey has performed more than 260 hours of observation on 39 different astronomical targets since December 2019. 

\section{The VERITAS  SII  (VSII) Observatory}
\begin{figure}[!h]
  \vspace{5mm}
  \centering
  \includegraphics[width=6in]{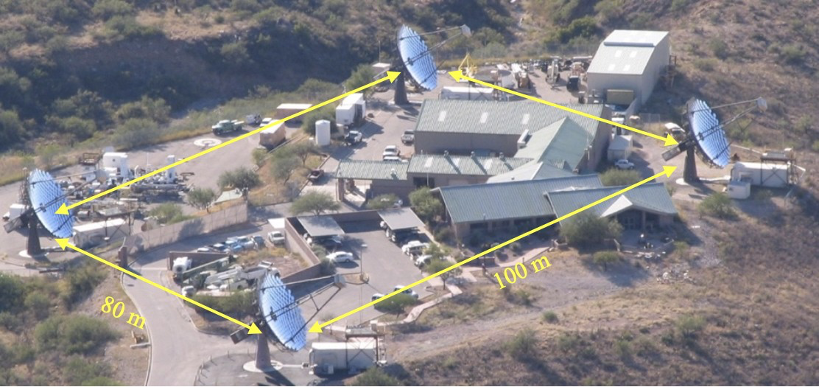}
  \caption{Telescope separations at the VERITAS gamma-ray observatory. The  physical baselines between VERITAS telescope pairs range between 80 to 175 meters. }
  \label{Figure1}
 \end{figure}

The VSII Observatory uses the optical telescopes of the VERITAS gamma-ray Observatory. The VERITAS Observatory  \cite{VERITAS} consists of an array of 4 IACTs located at an altitude of 1268 m.a.s.l at the F.L. Whipple Observatory near Amado, AZ. Each IACT contains an f/1.0 12 m-diameter segmented  Davies-Cotton reflector employing 345 hexagonal facets, resulting in 110 $m^2$ of light-collecting area. The lateral separation between nearest-neighbor IACT is approximately 80-120 m (Figure \ref{Figure1}). The Davies-Cotton reflector design creates approximately a 4 nanosecond spread to photons arriving at the telescope's focal plane. The combination of large telescope primary mirror area, fast optics, and 100+ meter telescope baselines enables VSII to provide sub-milliarcsecond angular resolution at short optical wavelengths (B/V bands)\cite{Matthews2020, SII2020}. 
 
\subsection{VSII Removable Focal Plane Instrumentation Plates}
\begin{figure}[!h]
  \vspace{5mm}
  \centering
  \includegraphics[width=5.in]{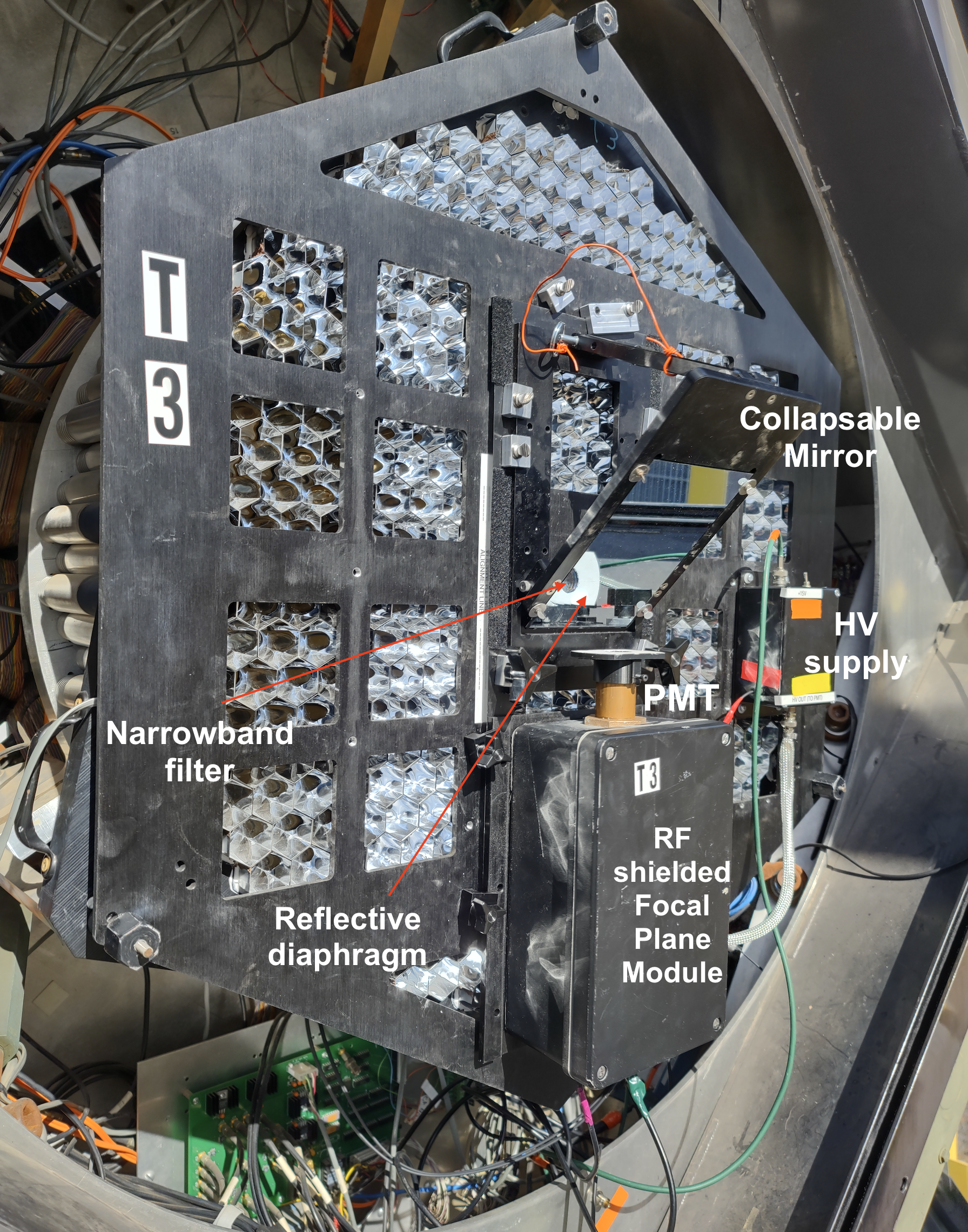}
  \caption{Veritas SII Focal Plane Plate (FPP) . Each removable FPP includes a collapsible mirror,  narrowband optical filter, PMT and preamplifier mounted within the RF shielded  Focal Plane Module, and a fiber optic controlled High Voltage (HV) supply.}
  \label{Figure2}
 \end{figure}

Each VERITAS telescope is instrumented with a removable VSII Focal Plane Plate (FPP)  that is securely mounted in front of the  499-pixel VERITAS camera (Figure \ref{Figure2}). 
  The removable FPPs  can be easily installed or removed in about 15 minutes per telescope by a single individual.  
The FPPs are designed to be securely mounted with no modifications or changes to the VERITAS cameras.
The FPP instrumentation employs a collapsible 45$^\circ$ mirror that reflects the starlight from the 12-m Davies-Cotton primary reflector onto an SII focal plane diaphragm that is oriented perpendicular to the telescope optical axis.
The collapsible mirror allows the VERITAS camera shutter to be closed during the daytime or bad weather.  This capability allows the SII FPPs to remain mounted on the VERITAS focal plane for the entire 5-10 day duration of the SII observing period.

The SII focal plane employs  a Semrock FF01-420/5-2 narrowband optical filter (420 nm/5 nm width)  \cite{semrock}  located in a diaphragm mount that is located directly in front of a fast light sensor (Hamamatsu super-bialkali R10560 photomultiplier tube (PMT)). The diaphragm mount includes a central aperture for the Semrock filter. A flat oval-shaped white reflective screen  enables the observer to center the light from the primary mirror onto the Semrock filter. The SII FPP instrumentation plate scale is automatically matched to the VERITAS focal plane point spread function (PSF) as it uses the same size PMTs used for the VERITAS gamma-ray camera.

\paragraph{Narrowband Optical Filter Effective Bandwidth}
The center bandpass wavelength of a narrowband optical interference filter will shift  to shorter wavelengths for light arriving at inclined angles.  The Semrock FF01-420 filter coating exhibits a high refractive index ($n = 2.38$) that reduces the wavelength shift for non-normal incident photons. A simulation was used to calculate the shift in optical bandpass wavelength for the VERITAS f/1.0 mirror optics by weighting the angular distribution of incoming photons by the relevant mirror area at each angle of incidence. The resulting incidence angle distribution was then convolved with the standard formula describing the shift in center bandpass wavelength with incidence angle\cite{semrock}. The resulting calculation indicates an effective bandwidth of $\delta \lambda \approx 13\ nm$  about a $\lambda_0 \approx 416\ nm$ center wavelength, with reduced light transmission compared to normal incidence.

\paragraph{Power supplies and signal transport}
 The PMT is powered by a custom battery-powered High Voltage  (HV) Power supply\cite{Cardon2019}. The HV supply battery uses two high-capacity (4400 ma-hr) Tenergy Li-Ion batteries permanently mounted in the VERITAS camera. The battery pack is recharged during the daytime.  When fully charged, the battery pack is sufficient to provide HV for several nights of SII observations. The HV setting is remotely controlled through a pulse width modulated  (PWM) signal supplied through a fiber optic interface. The fiber optic PWM signal is generated in the electronics trailer of each VERITAS telescope by an ethernet-controlled Arduino Yun connected to a fiber optic transmitter.   
 
The output of the PMT is amplified by a high-speed (200 MHz bandwidth) transimpedance ($2\times 10^4$ V/A) preamplifier (FEMTO HCA-200M-20K-C).  The distance between the PMT output and the input to the FEMTO preamplifier is kept as short as possible ($< $2 cm) to minimize the input capacitance load on the preamplifier input. This configuration maximizes the preamplifier bandwidth and decreases electronic noise.  The output from the FEMTO preamplifier drives a 45 m long double-shielded RG-223 coaxial cable. The cable is routed along the VERITAS telescope quadrapod arms and the optical support structure, and terminates at the SII data acquisition system, located in the telescope electronics trailer next to each VERITAS telescope. Special care is made to ensure that each SII focal plane instrumentation component is electrically isolated from any telescope ground to minimize ground-loop pickup of local RF noise. 

\begin{figure}[!h]
  \vspace{5mm}
  \centering
  \includegraphics[width=6.in]{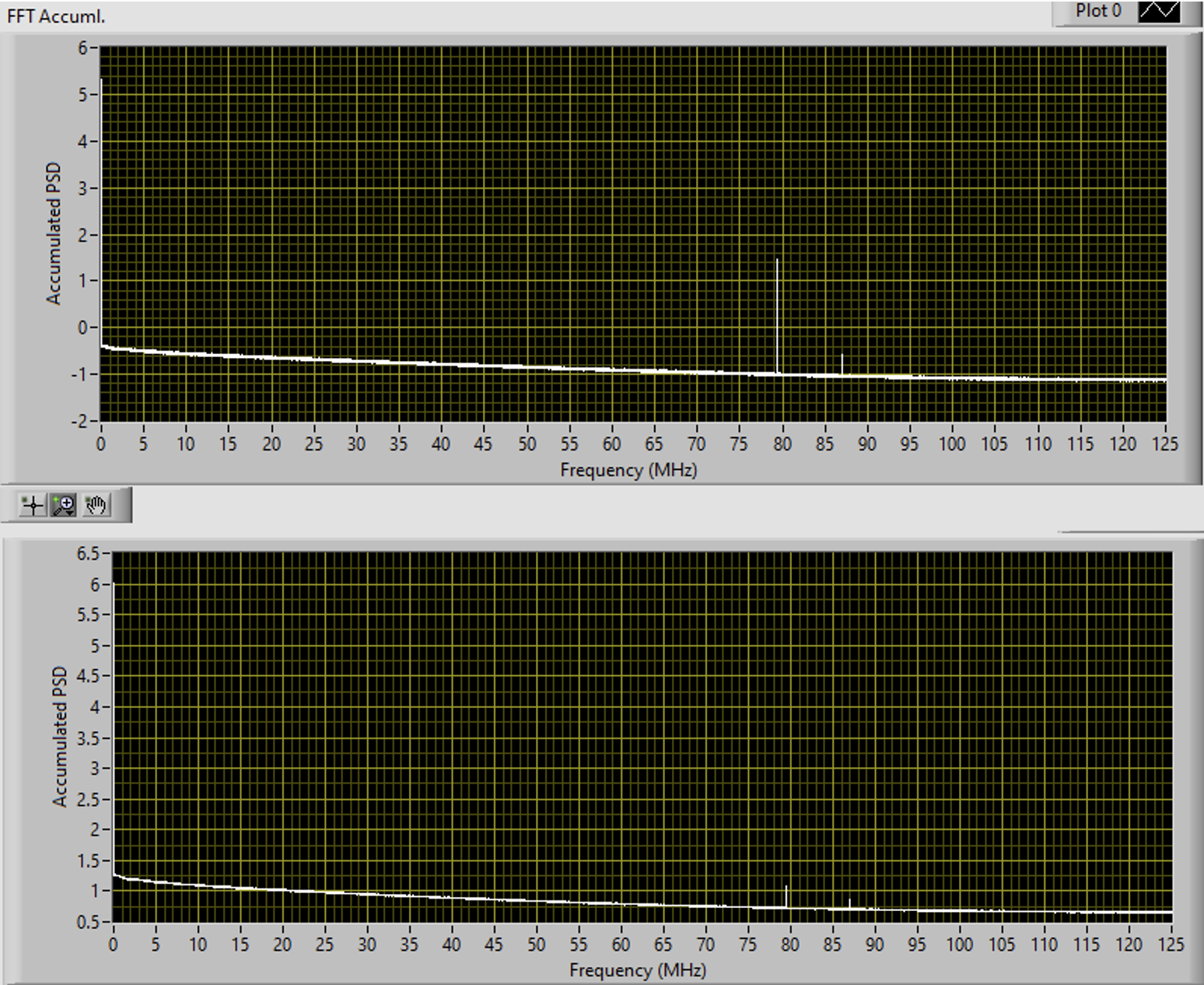}
  \caption{Observed RF noise in Telescope 3 focal plane instrumentation. The noise spectrum is calculated using a standard Fast Fourier Transform  (FFT) on raw data from Telescope 3  during an observing run. Horizontal Axis: Frequency (MHz). Vertical axis: FFT power spectrum (arbitrary units). Upper plot: Noise spectrum before RF shielding. Bottom Plot: Noise spectrum after RF shielding.}
  \label{Figure3}
 \end{figure}

\paragraph{RF Shielding}
The photomultiplier tube (PMT) is completely encased in a tight-fitting brass tube to reduce RF pickup into the PMT dynodes.  However, FFT analysis of data from each SII PMT in December 2020  revealed the presence of both persistent and transient RF noise (Figure \ref{Figure3}, upper panel). Prominent RF pickup specific to the FLWO observatory site is observed at a measured frequency of 79 MHz.  Additional RF shielding was added to the FPPs in January 2021 to reduce RF pickup. The PMT and the preamplifier were fully encased in an RF shielded enclosure (Focal Plane Module, or FPM). The FPM  reduced the RF pickup in each SII channel by a factor of 8 or more (Figure \ref{Figure3}, lower panel).  

\paragraph{Zero Baseline Beamsplitter FPP}
In June 2021, a specialized 'zero baseline (ZB) beamsplitter'  FPP was constructed and deployed on VERITAS Telescope 1.  The ZB FPP adds a non-polarizing 50-50 beamsplitter after the narrowband filter to the optical beam path. Two independent FPMs mounted perpendicular to each other on the FPP read out the split optical beam. Special care is taken to ensure the FPMs do not share a common ground connection on the FPP to eliminate ground-loop noise pickup.   Each FPM uses independent batteries, HV supplies, signal cables, and DAQ systems in the trailer.  Successful commissioning of the ZB FPP was performed using SII observations of several stars on the nights of June 24-26, 2021. 

\subsection{VSII Data Acquisition (DAQ) System}
A standalone National Instruments (NI) data acquisition system,  using a PXI eExpress (PXIe) crate with a high-speed backplane (4 GB/sec), is deployed at each VERITAS telescope to provide local data recording. Each SII data acquisition system. The crate holds a high-speed PXIe controller and a high-speed interface (PXIe-8262) connected to a high-speed (700 MB/sec) RAID disk (NI-8265). Each PMT signal is continuously digitized at 250 MHz by an NI FlexRio Module (NI-5761, 12-bit resolution,  DC-coupled). The 250 MHz digitization rate provides a sampling period that matches the 4 nsec time dispersion of the VERITAS optics. The digitized data from each PMT is truncated to 8 bits and merged into a single continuous data stream by a Virtex-5 FPGA processor. The data stream is transported across the high-speed PXIe backplane and recorded to the RAID disk. The VSII PXIe-based data acquisition uses commercially available modules, and the VSII data acquisition is controlled using LabView software.  

In Spring 2021, all VSII crate controllers were upgraded to include high-speed USB 3.0 ports (PXie-8135). Subsequently, we have demonstrated the ability to stream the SII  DAQ data directly onto USB 3.0 RAID disks. This capability provides a low-cost expansion of available disk space for nightly observations. As of July 2020, each SII DAQ system has more than 30 TB of high-speed data storage available for observations.  At a sampling rate of 250 Mhz (8-bit resolution), a one hour SII observation uses  865 GB of disk space.

\section{FPGA Clock Timing and  Telescope DACQ Synchronization} 
The intensity signals recorded between separated telescopes must be synchronized within a small fraction of the DAQ sampling period. If the DAQ clocks are not 100\% synchronized, the effective bandwidth of the measurement is degraded, resulting in a lower SIgnal-to-Noise Ratio (SNR). The distributed VSII data acquisition system incorporates synchronization of each telescope's DAQ FPGA clock using commercially available Seven Solutions White Rabbit  clock synchronization system\cite{WR}. The White Rabbit timing system uses single-mode fiber (SMF) optic links to distribute a centrally generated 10 MHz clock to each SII DAQ system telescope with a $< 200\ psec$ RMS precision. 

\paragraph{High-Speed Data Transport between VSII Telescopes}
An independent SII 10 GBase-T fiber optic data transfer network is used to provide high-speed network access to all four SII RAID data disks. This 10 Gb/sec bandwidth allows each VERITAS-SII DAQ system to cross-correlate its data stream with every other SII telescope data stream. The VSII  high-speed data transfer network is carried by 5/125 SMF  from each SII DAQ system to a central NetGear M4300-12X10 GBase-T managed ethernet switch.  The NetGear switch isolates the SII-related data traffic from the primary VERITAS observatory network.  An independent 1Gb/sec network interface is also established to each crate controller and HV control system in each telescope to allow access to these devices without competing for bandwidth with the main SII data stream. 

\subsection{FPGA-based Software Correlators}
Cross-correlation of the data streams between individual telescopes is performed using a  pipelined algorithm implemented on a National Instruments FPGA card mounted in the VSII DAQ systems.  The present correlation system searches the directory structure of the previous night's SII observations, matches individual telescope observations files into pairs for correlation, and then stages a batch process to perform all identified correlation pairs. A  1 hour SII observation generates 865 GB of data per telescope, and the FPGA correlator requires approximately 1.5 hours to complete the correlation of a single telescope pair.   The  VSII data transfer network has sufficient bandwidth to accommodate the simultaneous operation of two correlators on two independent DAQ systems.  For a one hour observation, approximately 4.5 hours of  VSII DAQ system use  is required to process all six baseline pairs.  Although some of the observations may be processed the next day, in general, an entire week of nightly SII observations requires an additional 7-10 days of computation to complete the correlations between every telescope pair.   

Each two-telescope correlation file is considerably smaller than the raw data file. A one-hour, two-telescope cross-correlation file is approximately 33 MB in size, compared to the combined  1.7 TB size of the two raw data files used to calculate the cross-correlation.  Once all two-telescope cross-correlations are completed, the correlation files are uploaded to the University of Utah Center for High Performance  Computing for archiving, distribution and analysis.  The majority of the raw data is discarded once the cross-correlations are completed. Secondary raw data files containing electronic calibrations and measurements of sky brightness are substantially smaller in size. Consequently,  it is possible to permanently retain them for correcting systematic effects for the final analysis.

\section{VSII Performance}
 VSII has been in regular 4-telescope operations on a  lunar-monthly (full-moon/rear-full-moon period) cadence since December 2019, recording more than 260 hours of observations to date.  VSII  was not in operation from March 2020 through September 2020 due to the COVID-related shutdown of the Whipple Observatory.   Additional discussion of the observing sequence,   target selection, and analysis,  including the observing constraints imposed by moon location and other considerations,  is found in the accompanying VSII survey paper \cite{Kieda2021}.  This paper also describes the analysis techniques for the VSII observations and the use of the resulting visibility curves to measure the angular diameter of nearby stars.
 
\section{Future Improvements}
In Fall 2021, the VERITAS-SII system will begin the installation of several additional improvements to increase VSII's operation efficiency and sensitivity. These improvements will include:
\begin{itemize}
\item Upgrade of RAID storage at each telescope to 100TB. This capacity will accommodate a full week of VSII observations at each telescope during the longest winter nights of the observing season.  
\item Development of improved narrowband optical filters. The updated filter will be custom designed to correct for the shift of bandpass frequency at non-zero incidence angles. When installed, these filters are expected to  increase Signal-to-Noise by a factor of 1.5-2.0.  
\item Automation of the telescope pointing corrections. The focal plane's CCD focal plane camera will be used to determine when a star is drifting from the SII photosensor. A machine learning algorithm will identify misalignments and send real-time corrections to each telescope's tracking position. 
\item Recoating of the VERITAS 12 m diameter primary mirror segments. The improved reflectivity will increase signal-to-noise by an additional factor of 1.5.
\item Improved RF shielding of the RG-233 Coaxial signal cables. The improved RF shielding will decrease RF pickup noise by a factor of 3 or more.  
\end{itemize}

\section{Acknowledgements}
This research is supported by grants from the US Department of Energy Office of Science, the US National Science Foundation, the Smithsonian Institution, and NSERC in Canada. This research used resources provided by the Open Science Grid, which is supported by the National Science Foundation and the US Department of Energy's Office of Science, and resources of the National Energy Research Scientific Computing Center (NERSC), a US Department of Energy Office of Science User Facility operated under Contract No. DE-AC02-05CH11231. The authors gratefully acknowledge support under NSF Grant \#AST 1806262 for the fabrication and commissioning of the VERITAS-SII instrumentation. We acknowledge the excellent work of the technical support staff at the Fred Lawrence Whipple Observatory and the collaborating institutions in the construction and operation of the instrument.

%% Full authors list (ONLY FOR COLLABORATIONS)

 \clearpage
 \section*{Full Authors List: \Coll\ Collaboration}
 
 \scriptsize
 \noindent
 C.~B.~Adams$^{1}$,
 A.~Archer$^{2}$,
 W.~Benbow$^{3}$,
 A.~Brill$^{1}$,
 J.~H.~Buckley$^{4}$,
 M.~Capasso$^{5}$,
 J.~L.~Christiansen$^{6}$,
 A.~J.~Chromey$^{7}$, 
 M.~Errando$^{4}$,
 A.~Falcone$^{8}$,
 K.~A.~Farrell$^{9}$,
 Q.~Feng$^{5}$,
 G.~M.~Foote$^{10}$,
 L.~Fortson$^{11}$,
 A.~Furniss$^{12}$,
 A.~Gent$^{13}$,
 G.~H.~Gillanders$^{14}$,
 C.~Giuri$^{15}$,
 O.~Gueta$^{15}$,
 D.~Hanna$^{16}$,
 O.~Hervet$^{17}$,
 J.~Holder$^{10}$,
 B.~Hona$^{18}$,
 T.~B.~Humensky$^{1}$,
 W.~Jin$^{19}$,
 P.~Kaaret$^{20}$,
 M.~Kertzman$^{2}$,
 D.~Kieda$^{18}$,
 T.~K.~Kleiner$^{15}$,
 S.~Kumar$^{16}$,
 M.~J.~Lang$^{14}$,
 M.~Lundy$^{16}$,
 G.~Maier$^{15}$,
 C.~E~McGrath$^{9}$,
 P.~Moriarty$^{14}$,
 R.~Mukherjee$^{5}$,
 D.~Nieto$^{21}$,
 M.~Nievas-Rosillo$^{15}$,
 S.~O'Brien$^{16}$,
 R.~A.~Ong$^{22}$,
 A.~N.~Otte$^{13}$,
 S.~R. Patel$^{15}$,
 S.~Patel$^{20}$,
 K.~Pfrang$^{15}$,
 M.~Pohl$^{23,15}$,
 R.~R.~Prado$^{15}$,
 E.~Pueschel$^{15}$,
 J.~Quinn$^{9}$,
 K.~Ragan$^{16}$,
 P.~T.~Reynolds$^{24}$,
 D.~Ribeiro$^{1}$,
 E.~Roache$^{3}$,
 J.~L.~Ryan$^{22}$,
 I.~Sadeh$^{15}$,
 M.~Santander$^{19}$,
 G.~H.~Sembroski$^{25}$,
 R.~Shang$^{22}$,
 D.~Tak$^{15}$,
 V.~V.~Vassiliev$^{22}$,
 A.~Weinstein$^{7}$,
 D.~A.~Williams$^{17}$,
 and 
 T.~J.~Williamson$^{10}$\\
 \noindent
 $^{1}${Physics Department, Columbia University, New York, NY 10027, USA}
 $^{2}${Department of Physics and Astronomy, DePauw University, Greencastle, IN 46135-0037, USA}
 $^{3}${Center for Astrophysics $|$ Harvard \& Smithsonian, Cambridge, MA 02138, USA}
 $^{4}${Department of Physics, Washington University, St. Louis, MO 63130, USA}
 $^{5}${Department of Physics and Astronomy, Barnard College, Columbia University, NY 10027, USA}
 $^{6}${Physics Department, California Polytechnic State University, San Luis Obispo, CA 94307, USA} 
 $^{7}${Department of Physics and Astronomy, Iowa State University, Ames, IA 50011, USA}
 $^{8}${Department of Astronomy and Astrophysics, 525 Davey Lab, Pennsylvania State University, University Park, PA 16802, USA}
 $^{9}${School of Physics, University College Dublin, Belfield, Dublin 4, Ireland}
 $^{10}${Department of Physics and Astronomy and the Bartol Research Institute, University of Delaware, Newark, DE 19716, USA}
 $^{11}${School of Physics and Astronomy, University of Minnesota, Minneapolis, MN 55455, USA}
 $^{12}${Department of Physics, California State University - East Bay, Hayward, CA 94542, USA}
 $^{13}${School of Physics and Center for Relativistic Astrophysics, Georgia Institute of Technology, 837 State Street NW, Atlanta, GA 30332-0430}
 $^{14}${School of Physics, National University of Ireland Galway, University Road, Galway, Ireland}
 $^{15}${DESY, Platanenallee 6, 15738 Zeuthen, Germany}
 $^{16}${Physics Department, McGill University, Montreal, QC H3A 2T8, Canada}
 $^{17}${Santa Cruz Institute for Particle Physics and Department of Physics, University of California, Santa Cruz, CA 95064, USA}
 $^{18}${Department of Physics and Astronomy, University of Utah, Salt Lake City, UT 84112, USA}
 $^{19}${Department of Physics and Astronomy, University of Alabama, Tuscaloosa, AL 35487, USA}
 $^{20}${Department of Physics and Astronomy, University of Iowa, Van Allen Hall, Iowa City, IA 52242, USA}
 $^{21}${Institute of Particle and Cosmos Physics, Universidad Complutense de Madrid, 28040 Madrid, Spain}
 $^{22}${Department of Physics and Astronomy, University of California, Los Angeles, CA 90095, USA}
 $^{23}${Institute of Physics and Astronomy, University of Potsdam, 14476 Potsdam-Golm, Germany}
 $^{24}${Department of Physical Sciences, Munster Technological University, Bishopstown, Cork, T12 P928, Ireland}
 $^{25}${Department of Physics and Astronomy, Purdue University, West Lafayette, IN 47907, USA}

 \end{document}